\documentclass{pnastwo}

\usepackage{amssymb,amsfonts,amsmath}
\usepackage{graphics,bm}
\usepackage{epsfig}
\usepackage{epsf}

\url{www.pnas.org/cgi/doi/10.1073/pnas.0709640104}
\copyrightyear{2008}
\issuedate{Issue Date}
\volume{Volume}
\issuenumber{Issue Number}

\def\be{\begin{equation}} \def\ee{\end{equation}}
\def\beq{\begin{eqnarray}} \def\eeq{\end{eqnarray}}

\def\nn{\nonumber}

\begin{document}

\title{Theory of Point Contact Spectroscopy in Correlated Materials}

\author{Wei-Cheng Lee\affil{1}{Department of Physics, University of Illinois, 1110 West Green Street, Urbana, Illinois 61801, USA},
Wan Kyu Park\affil{1}{},
Hamood Z. Arham\affil{1}{},
Laura H. Greene\affil{1}{},
\and
Philip W. Phillips\affil{1}{}}

\contributor{Submitted to Proceedings of the National Academy of Sciences
of the United States of America}

\maketitle

\begin{article}
\begin{abstract}
{We develop a microscopic theory for the point-contact conductance between a metallic electrode and a strongly correlated material using the non-equilibrium Schwinger-Kadanoff-Baym-Keldysh formalism. We explicitly show that in the classical limit, contact size shorter than the scattering 
length of the system, the microscopic model can be reduced to an effective model with transfer matrix elements that conserve in-plane momentum.
We find that the conductance $dI/dV$ is proportional to the {\it effective density of states}, that is, the integrated single-particle spectral function $A(\omega=eV)$ over the 
whole Brillouin zone.  From this conclusion, we are able to establish
the conditions under which a non-Fermi liquid metal exhibits a zero-bias
peak in the conductance.  This finding is discussed in the context of
recent point-contact spectroscopy on the iron pnictides and chalcogenides which
has exhibited a zero-bias conductance peak.}
\end{abstract}

\section{Significance Statement}
Point-contact spectroscopy is a bulk spectroscopic probe that has been reliably used to map out bosonic and superconducting order parameter spectra via quasiparticle classical 
and Andreev scattering, respectively.  We had previously shown this technique to be exquisitely sensitive to an {\it effective} density of states specifically arising from 
non-Fermi liquid behavior, and in the case of the iron pnictides and chalcogenides, electronic nematicity manifesting as a zero bias conductance peak corresponds to an increased 
effective density of states at the Fermi level arising from orbital fluctuations. We have developed a quantum mechanical theory to show how this technique reveals such 
effective density of states, while being insensitive to gapless Fermi surface reconstructions; and is therefore a valuable filter for detecting non-Fermi liquid behavior.
\section{Introduction}
Heavy fermion systems\cite{hfreview1,hfreview2}, high-$T_c$ cuprates\cite{leecupratereview,phillipsreview}, and very recently 
the iron-based superconductors\cite{daireview,leewcreview} all exhibit
symptoms of quantum criticality. The most striking feature of  quantum criticality is that the quantum fluctuations associated with the 
quantum critial point (QCP) couple strongly to itinerant
electrons giving rise to drastic changes in the electronic
properties. Typically, such emergent properties are non-Fermi
liquid-like and hence fall outside the standard theory of metals.
While measurements of several physical properties, for example, the heat capacity, magnetic susceptibility, and DC electrical resistivity, have been identified with 
NFL behavior, a direct probe of the hallmark feature of a NFL, namely the imaginary part of the single-particle self energy $\Sigma(\omega)\sim \omega^\nu$ with $\nu < 1$, 
is still lacking.   
In principle, the temperature dependence of the DC electrical
resistivity is expected to be related to $\nu$, but it is also
sensitive to many other factors, rendering such measurements inconclusive. In this context angle-resolved photoemission (ARPES)
is an ideal probe of this hallmark feature. However, the resolution of
the ARPES data is typically not high enough to pin-down $\nu$ conclusively.
As a result, a reliable experimental setup to judge whether $\nu$ is larger or smaller than 1 is one of the most important topics in this field.

We demonstrate here how point contact spectroscopy (PCS) can be used to
resolve this problem.   Our work here is motivated by recent PCS experiments on iron-pnictide superconductors in which an excess zero-bias conductance was measured well above the temperature associated with the structural rearrangement.  Based on an analogy with earlier theoretical work on nematic quantum phase transitions\cite{lawler2006},  Lee, et al. argued that the excess zero-bias conductance measured experimentally is likely due to an excess density of states associated with fluctuations near the orbital-ordering quantum phase transition.  However, a direct link between the two remains missing as there is no rigorous argument relating the PCS signal in strongly correlated systems to the single-particle density of states. 
The theoretical foundations for the tunneling density of states have been well established\cite{caroli,tersoff}, but we stress here that PCS is not tunneling, 
and this work establishes a clear connection between PCS conductance and an effective density of states arising from NFL behavior.

In this paper, we build on earlier work\cite{fogelstr2010} to fill in this missing link and as a result are able to establish the circumstances under which the PCS signal is a direct measure of the single-particle density of states in strongly correlated electron systems and hence offers a window into a key probe of non-Fermi liquid behavior, namely the imaginary part of the single-particle self energy. Of course point-contact spectroscopy is an old field dating back to the pioneering work of  I. K. Yanson's \cite{yanson1974} in 1974 when he was attempting to measure the tunneling conductance across a 
superconducting/insulator/normal-metal (SIN) planar junction.  According to Harrison's theorem \cite{harrison1961}, when the superconductor is driven normal, 
the resulting planar tunneling conductance must be ohmic:  Assuming a one-dimensional model, which holds for good planar tunnel junctions, in the 
standard tunneling conductance formula that assumes weakly correlated electron (conventional) materials, the Fermi velocity $v_f=(1/\hbar)(dk_x/dE)$  exactly divides out the 
density of states $D(E) = (L/\pi)(dE/dk_x)$. Yanson discovered weak conductance non-linearities in an SIN junction above the critical field (S = Pb), and 
that the second harmonic of the conductance $(d^2 I/dV^2)$ revealed the Eliashberg function, $\alpha^2F(\omega)$ the strength of about $1\%$ of the background conductance.  
This resulted from the junctions being "leaky" with nanoscale metallic shorts between S and N allowing electrons to be directly injected through the junction without any 
tunneling processes; hence Harrison's theorem did not apply. He went on to show that when the junction is large such that the mean free path is smaller than the junction 
(thermal regime) no spectroscopic information can be obtained, but if the junction is smaller than the elastic (Sharvin or ballistic limit) or inelastic (diffusive regime) 
mean free path, spectroscopic information is revealed. 
Quasiparticles backscattered through the junction reveal the phonon spectrum; so this technique is also called quasiparticle scattering spectroscopy (QPS).  
Researchers in the field proceeded to map out bosonic spectra (phonons and magnons) in a variety of materials \cite{jansen1980,khotkevich1995}, 
quasiparticle scattering from Kondo impurities \cite{jansen1980}, spin and charge density waves \cite{meekes1988, escudero2010}, and 
recently a Kondo insulator\cite{Zhang2013}.  With the advent of the Blonder-Tinkham-Klapwijk (BTK) theory\cite{btk1982}, in a clean S/N junction, 
Andreev scattering was shown to reveal details of the superconducting gap structure, including magnitude and symmetry\cite{park2012}.  
This technique has been shown to be particularly useful in superconductors that are difficult to grow in thin film form, e.g., 
heavy-fermion superconductors \cite{goll2005, park2008, fogelstr2010} and the iron-based high-temperature superconductors \cite{daghero2011}.  
In the heavy-fermions, the Fano background could be accounted for via multichannel tunneling models\cite{fogelstr2010,yang2008,haule2009,maltseva2009}. 
Our focus here is on establishing a clear link between the suggestive relationship that  excess zero-bias signal measured in PCS is a direct measure of the {\it effective density of states} arising from electron correlations
\cite{leewcreview,park2008, park2012, arham2012, leewc2012nfl}.

To this end,  we use the Schwinger-Kadanoff-Baym-Keldysh (SKBK) formalism\cite{schwinger,kadanoff-baym,lvkeldysh}, coupled with certain
quite reasonable assumptions, to show that the conductance measured from PCS is proportional to the {\it effective} density of states\cite{noting}. 
Since the effective density of states is defined as the integrated single-particle spectral function $A(\omega=eV)$ over the
whole Brillouin zone, it contains the information about the single
particle self energy $\Sigma(\omega)$. We show that a fingerprint of 
 non-Fermi liqud behaviour with $\nu<1$ is an
enhancement of the PCS conductance at zero bias. 
As a comparison, we also discuss the case of a junction in the thermal regime, in which only DC resistivity is detected. 
We highlight here that the DC resistivity is fundamentally different from the PCS conductance. 
The former is corresponding to the {\it current-current} correlation function evaluated by Kubo formula, while the later is related to the single particle Green function
as described by the SKBK formalism.
Consequently, we conclude that the zero-bias peak in the PCS could be identified as a 
unique signature of non-Fermi liquid metal.


\section{Hamiltonian}
\begin{figure}
\includegraphics{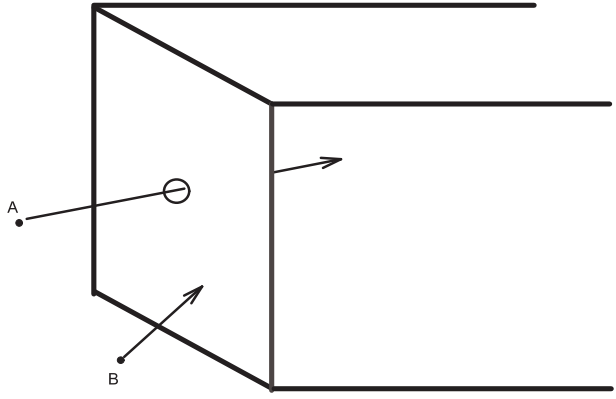}
\caption{\label{fig:pcsgeometry} Ideal geometry of point-contact spectroscopy. Electron A can transport from the metallic electrode into the system through the contact (circle), thereby
contributing to the current, while electron B is scattered back by the
wall, resulting in no contribution to the current.}
\end{figure}

We start from the ideal point-contact geometry shown in
Fig. \ref{fig:pcsgeometry}. The left-hand side is a metallic electrode with infinite bandwidth, and electrons 
can transport from the electrode into the system (right-hand side) via point contact. 
A general model to describe this geometry is given by
\beq
H&=&H_{electrode}+H_c+H_{sys}\nn\\
H_{electrode} &=& \sum_{\vec{k},\sigma,\alpha\in L,R} \epsilon_{\vec{k},\alpha} d^\dagger_{\vec{k},\sigma,\alpha}d_{\vec{k},\sigma,\alpha},\nn\\
H_c &=& \sum_{\vec{k},\sigma,\alpha\in L,R,\vec{p},\sigma'} V^{\sigma,\sigma'}_{\vec{k}.\alpha,\vec{p}} d^\dagger_{\vec{k},\sigma,\alpha} c_{\vec{p},\sigma'} + h.c.,\nn\\
H_{sys} &=& \sum_{\vec{k},\sigma} E(\vec{k}) c^\dagger_{\vec{k},\sigma} c_{\vec{k},\sigma} + H_I,
\label{ham}
\eeq 
where $d^\dagger_{\vec{k},\sigma,\alpha}$ refers to the creation operator for an electron with momentum $\vec{k}$ and spin $\sigma$ in the right ($\alpha=R$) or left ($\alpha=L$) 
electrode, $c^\dagger_{\vec{k},\sigma}$ creates an electron with
momentum $\vec{k}$ and spin $\sigma$ in the system and
$\epsilon_{\vec{k},\alpha}$ is the band-structure energy dispersion for the electrodes.
$H_{sys}$ is the total Hamiltonian for the system to be measured, containing the kinetic energy with band dispersion of $E(\vec{k})$ and the interaction Hamiltonian $H_I$. 
$H_c$ describes the transfer of electrons between the electrode and the system via point contact.

To determine the $H_c$ suitable for a realistic point-contact geometry, we first write down the corresponding term in real space as
\be
H_c = \sum_{\sigma,\sigma'} \int d\vec{r} V^{\sigma,\sigma'}(\vec{r}) \psi^\dagger_{d,\sigma}(\vec{r}) \psi_{c,\sigma'}(\vec{r}) + h.c.,
\label{hc1}
\ee
where $\psi^\dagger_{d,c}(\vec{r})$ are the creation operators for
electron in the electrode and system. $V^{\sigma,\sigma'}(\vec{r})$
which can describe the point contact shown in 
Fig. \ref{fig:pcsgeometry} should be written as
\beq
V^{\sigma,\sigma'}(\vec{r}) &=& \delta_{\sigma,\sigma'}V\,\,\,,\,\,\,\vec{r}\in {\rm contact}\nn\\
&=& 0\,\,\,,\,\,\,{\rm otherwise}.
\label{vr}
\eeq
In Eq. (\ref{vr}), we have assumed that the spin is conserved through the contact. Performing a Fourier transformation on Eq. (\ref{hc1}) leads to $H_c$ given in Eq. (\ref{ham}). 

Although the exact form of $V^{\sigma,\sigma'}_{\vec{k}.\alpha,\vec{p}}$ depends on the experimental setup, a general structure for it can still be obtained.
From the uncertainty principle, if the electron wave function were to be squeezed into a small area described by Eq. (\ref{vr}), 
the uncertainty in the momentum of the electron would increase. 
In other words, as the electron goes through the contact, it could change momentum due to the uncertainty principle. As a result, $V_{\vec{k},\alpha,\vec{p}}$ is non-zero 
in general for $\vec{k}\neq \vec{p}$. Furthermore, if the contact size
were to increase (larger uncertainty in position), the uncertainty in
momentum would decrease. Therefore, 
$V_{\vec{k},\alpha,\vec{p}}$ for $\vec{k}\neq \vec{p}$ becomes smaller and smaller as the contact size increases. 
We consider two extreme cases. First, if $V(\vec{r}) = \delta(\vec{r}-\vec{r}_0)$, which corresponds to the case for STM, $V_{\vec{k},\alpha,\vec{p}} = 1$. 
This means that an electron with momentum $\vec{k}$ has an equal
probability of changing  momentum to $\vec{p}$ after going through the delta-function point contact. 
On the other hand, for the planar tunnel junction (large contact regime), we assume that the matrix elements of electrons tunneling throughout 
the entire interface are the same; that is, $V(\vec{r}) = V$ for $r_z =
0$. In this case, since the uncertainty in the position of the
interface is infinite, the uncertainty in the in-plane 
momentum is zero. Therefore we have $V_{\vec{k},\alpha,\vec{p}} \neq 0$ only if the in-plane momentum is conserved ($\vec{k}_\parallel=\vec{p}_\parallel$). 
This gives us the well-known tunneling model.

\begin{figure}
\includegraphics{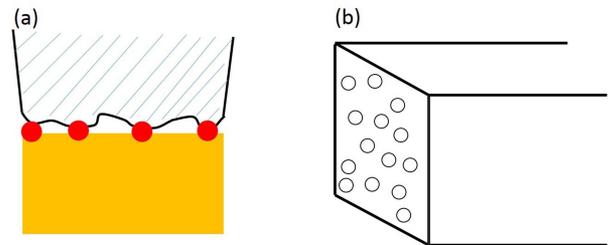}
\caption{\label{fig:realpcs} (a) Cartoon representation for the realistic point contact junction. A metal (shaded area) is coated with an insulating material (bold line), and
several shorts (red dots) are in contact with the material to be measured. (b) The corresponding picture of the interface with randomly distributed point contacts.}
\end{figure}

In terms of momentum conservation, PCS lies between the two extremes of STM and planar tunneling.  In PCS, the in-plane momentum is not necessarily conserved due to the quantum effects.
However, if the contact size is large enough but still smaller than the electron mean-free path, $V_{\vec{k},\alpha,\vec{p}}$ is peaked at $\vec{k}_\parallel = \vec{p}_\parallel$. 
Therefore the classical scatterings, meaning scattering events conserving in-plane momentum, will be dominating.
Moreover, experimentally a point contact junction is comprised of
many randomly distributed short areas as demonstrated in Fig. \ref{fig:realpcs}(a).
Such a configuration with many randomly distributed point contacts (shown in Fig. \ref{fig:realpcs}(b)) allows electrons pass through the junction via these different locations. 
This effectively increases the uncertainty in position and leads to a further suppression of matrix elements 
$V_{\vec{k},\alpha,\vec{p}}$ for $\vec{k}_\parallel \neq \vec{p}_\parallel$, as long as the average distance between these point contacts are comparable to, or smaller than, the 
electron mean-free path.
Consequently, it is a valid approximation that the dominant scatterings
are ballistic, and in keeping with the classical limit,  we  keep only
that part of the transfer Hamiltonian which conserves the in-plane momentum.
This reduces our model to that of an effective Hamiltonian containing only transfer matrix elements that conserve the in-plane momentum, which has
been shown to capture the physics for PCS in the classical limit in several systems\cite{maltseva2009,fogelstr2010}. 

\section{Schwinger-Kadanoff-Baym-Keldysh (SKBK) formalism}
Now we briefly discuss the working principle of PCS. Electrons are injected from the metallic electrode (left) via the point contact into the system with excess energy. 
This excess energy must be relaxed by inelastic scattering in the
system, and the electrons must be thermalized upon reaching the
electrode on the other side (right), thereby
contributing to the current $I$ that can be measured.
Consequently, the conductance, defined as $(dI/dV)$, could reveal
information about scattering mechanisms in the system.
Since the processes involving electrons relaxing excess energy are non-equilibrium in nature,
we employ the standard SKBK formalism to compute the current.  
By contrast, if the contact size is longer than the electron mean-free path, i.e., in the thermal regime, the excess energy of the injected electrons will be consumed at the junction. 
In this case, the SKBK formalism does not apply and one should treat it as a usual resistor using the Kubo formula. 
As a result, in the thermal regime, PCS conductance will only be determined by the resistivity and in fact, this criterion has been used to determine the quality of the junction in 
Ref. \cite{arham2012}.

We follow the same procedure and notation used in Ref. \cite{keldysh} for the SKBK formalism.
The final expression for the current is
\be
I = \frac{e}{\hbar} \int \frac{d\epsilon}{2\pi} {\rm Tr}\left[\frac{\hat{\Gamma}^L(\epsilon)\hat{\Gamma}^R(\epsilon)}{\hat{\Gamma}^L(\epsilon) + \hat{\Gamma}^R(\epsilon)} 
\hat{A}(\epsilon)\right]\times\left[f_L(\epsilon)-f_R(\epsilon)\right],
\label{finalI}
\ee
where $f_{L,R}(\epsilon)$ is the Fermi-Dirac function of left (right) electrodes, and
\be
\hat{A}(\epsilon) = i\left[\hat{G}^r(\epsilon) - \hat{G}^a(\epsilon)\right]
\label{jl}
\ee
is the single-particle spectral function defined as the imaginary part of the full Green function. $\hat{G}^{r,a}(\epsilon)$ are the retarded and advanced Green functions containing 
self-energies
\be
\hat{\Sigma}(\epsilon) = \hat{\Sigma}_{sys}(\epsilon) + \hat{\Sigma}_L(\epsilon) + \hat{\Sigma}_R(\epsilon),
\label{selfall}
\ee
with $\hat{\Sigma}_{sys}(\epsilon)$ the self-energy due to the interactions inside the system $H_I$ and $\hat{\Sigma}_{L,R}(\epsilon)$ the self-energy due to the contacts.
$\hat{\Gamma}^{L,R}(\epsilon)$ is the imaginary part of $\hat{\Sigma}_{L,R}(\epsilon)$. All the functions here are matrices with indices $(m,n)$ standing for the degrees of freedom in $H_{sys}$.

It is instructive to analyze certain limits of Eq. (\ref{finalI}).  As argued above, the matrix element $V^{\sigma,\sigma'}_{\vec{k},\alpha,\vec{p}}$ can be well approximated
by
\be
V^{\sigma,\sigma'}_{\vec{k},\alpha,\vec{p}} = V_{\vec{p}_\parallel} \delta_{\sigma,\sigma'} \delta_{\vec{k}_\parallel,\vec{p}_\parallel}.
\label{vkp}
\ee
With this choice of $V^{\sigma,\sigma'}_{\vec{k},\alpha,\vec{p}}$, we find that $\hat{\Gamma}^{L,R}(\epsilon)$ is diagonal in momentum space. 
As a result, if $\hat{\Sigma}_{sys}(\epsilon)$ is also diagonal in momentum space, which is usually the case for states not breaking translational invariance, 
Eq. (\ref{finalI}) becomes
\beq
I &=& \frac{e}{\hbar} \int d\vec{p}_\parallel\int \frac{d\epsilon}{2\pi} \left[\frac{\Gamma^L(\vec{p}_\parallel,\epsilon)\Gamma^R(\vec{p}_\parallel,\epsilon)}
{\Gamma^L(\vec{p}_\parallel,\epsilon) + \Gamma^R(\vec{p}_\parallel,\epsilon)}\right]
A(\vec{p}_\parallel,\epsilon)\nn\\
&\times&\left[f_L(\epsilon)-f_R(\epsilon)\right].
\label{finalI2}
\eeq
If we assume $\hat{\Gamma}^{L,R}(\vec{p}_\parallel,\epsilon)$ lacks
momentum dependence and work in the limit of ${\rm Im}\Sigma$
small enough so that
$A(\vec{p}_\parallel,\epsilon)$ becomes a delta-function,
Eq. (\ref{finalI2}) reduces to
\be
I = \frac{e}{\hbar} \int \frac{d\epsilon}{2\pi} T(\epsilon) D(\epsilon)\left[f_L(\epsilon)-f_R(\epsilon)\right],
\ee
where the density of states for the quantum channel is
\be
D(\epsilon) = \int d\vec{p}_\parallel
\delta(\epsilon-E_{\vec{p}_\parallel})
\label{dos}
\ee
and 
\beq
T(\epsilon) = \left[\frac{\Gamma^L(\epsilon)\Gamma^R(\epsilon)}
{\Gamma^L(\epsilon) + \Gamma^R(\epsilon)}\right]
\eeq
is the transmission coefficient.
This is of the same form of the Landauer transport formula.

\section{Formalism for conductance}

Since the conductance is obtained by taking the derivative of the current with respect to the applied voltage, $V$, it is crucial to find out the $V$-dependence in the expression 
for $I$.   We continue using the approximation given in Eq. (\ref{vkp}) and consequently adopt the expression for $I$ given in Eq. (\ref{finalI2}). 
As usual, the Fermi functions $\left[f_L(\epsilon)-f_R(\epsilon)\right]$ give rise to the most prominent $V$-dependence. 
We make the common assumptions that the electrodes have infinite
bandwidth and the chemical potentials for the two electrodes are
related through 
$eV = \mu_L-\mu_R$. Without loss of generality, we set $\mu_R=0$ and
consequently, $\mu_L=eV$.  The effect of the external electric field
due to the voltage bias on the energy dispersion of the system has not
been included at present. 
This effect is generally complicated, but it can be included phenomenologically in our formalism by replacing $E_{\vec{p}_\parallel}$ with $E_{\vec{p}_\parallel}(eV)$.
As a consequence, at zero-temperature, the current $I$ becomes
\be
I = \frac{e}{\hbar} \int d\vec{p}_\parallel\int^{eV}_0 \frac{d\epsilon}{2\pi} T(\vec{p}_\parallel,\epsilon)
A(\vec{p}_\parallel,eV,\epsilon),
\label{finalI3}
\ee
where
\beq
&&T(\vec{p}_\parallel,\epsilon)=  \left[\frac{\Gamma^L(\vec{p}_\parallel,\epsilon)\Gamma^R(\vec{p}_\parallel,\epsilon)}
{\Gamma^L(\vec{p}_\parallel,\epsilon) + \Gamma^R(\vec{p}_\parallel,\epsilon)}\right],\nn\\
&&A(\vec{p}_\parallel,eV,\epsilon)\nn\\
&=& \frac{{\rm Im}\Sigma(\vec{p}_\parallel,\epsilon)}{\big[\epsilon - (E_{\vec{p}_\parallel}(eV)-\mu) - 
{\rm Re}\Sigma(\vec{p}_\parallel,\epsilon)\big]^2 + \left[{\rm Im}\Sigma(\vec{p}_\parallel,\epsilon)/2\right]^2}.\nn\\
\label{apw}
\eeq

So far the formalism is still general, but more realistic approximations can be made to simplify the calculations. 
If the interface between the electrodes and the system is clean, no extra scatterings will be induced by the surface effects.
Therefore, we can further assume $V_{\vec{p}_\parallel} = V_0$ in Eq. (\ref{vkp})\cite{noteAR}.
With this assumption in hand, we reduce the self energy due to the contact to $\Sigma_{L,R}(\vec{p}_{\parallel},\epsilon)\sim \Lambda^{L,R} + i \Gamma^{L,R}$,
where $\mathcal{N}$ is the density of states near the Fermi energy $E_f$ of the electrodes. 
The last simplification in the above equation reflects the fact that we are only interested in the energy scale $eV\sim 100meV << E_f$.
In other words, $T(\vec{p}_\parallel,\epsilon)$ can be approximated by a constant $T=\Gamma^L\Gamma^R/(\Gamma^L+\Gamma^R)$, 
and the conductance can then be obtained by
\beq
\frac{dI}{dV} &=& \mathcal{G}_0(eV) + \mathcal{G}_1(eV),\nn\\
\mathcal{G}_0(eV) &=&  \frac{e^2 T}{h} \int d\vec{p}_\parallel 
A(\vec{p}_\parallel,eV,\epsilon=eV),\nn\\
\mathcal{G}_1(eV) &=&\frac{eT}{\hbar} \int d\vec{p}_\parallel\int^{eV}_0 \frac{d\epsilon}{2\pi}
\frac{d A(\vec{p}_\parallel,eV,\epsilon)}{dV}.
\eeq

It is now clear that the most dominant term in the conductance is
$\mathcal{G}_0(eV)$ which encodes the spectral function.
$\mathcal{G}_1(eV)$ describes the contribution due to the effect of external electric field to the system. 
Such an effect could be generally small except in systems like ferroelectric and nano materials. 
Moreover, since PCS is a bulk measurement, it is reasonable to
assume that $E_{\vec{p}_\parallel}(eV) = E_{\vec{p}_\parallel} - eV/2$, meaning that 
the bias voltage only produces a shift in the energy dispersion.
For $eV<<E_f$, it is clear that the change in the spectral function due to $eV$, equivalently $\frac{d A(\vec{p}_\parallel,eV,\epsilon)}{dV}$ in 
$\mathcal{G}_1(eV)$, is negligible for our purpose.

Collecting all the pieces together, we obtain
\be
\frac{dI}{dV} = \frac{e^2 T}{h} \int d\vec{p}_\parallel \mathcal{A}(\vec{p}_\parallel,eV),
\ee
as our final expression for the conductance, where
\beq
&&\mathcal{A}(\vec{p}_\parallel,eV)\nn\\
&=& \frac{{\rm Im}\Sigma(\vec{p}_\parallel,\epsilon)}{\big[\frac{3}{2}eV - (E_{\vec{p}_\parallel}-\mu) -
{\rm Re}\Sigma(\vec{p}_\parallel,\epsilon)\big]^2 + \left[{\rm Im}\Sigma(\vec{p}_\parallel,\epsilon)\right]^2}.\nn\\
\label{finalG}
\eeq

\section{Application to iron pnictides}
Eq. \ref{finalG} allows us to compute the PCS conductance for a variety of systems as long as the single particle self-energy is known. 
It can be shown that for the non-interacting case, PCS conductance actually detects the band structure density of states. 
For a non-Fermi liquid system, a unique zero-bias peak will be obtained if the single particle self energy satisfies some criteria. 
The proof is straightforward, as detailed in the Supplementary Materials.

\begin{figure}
\includegraphics{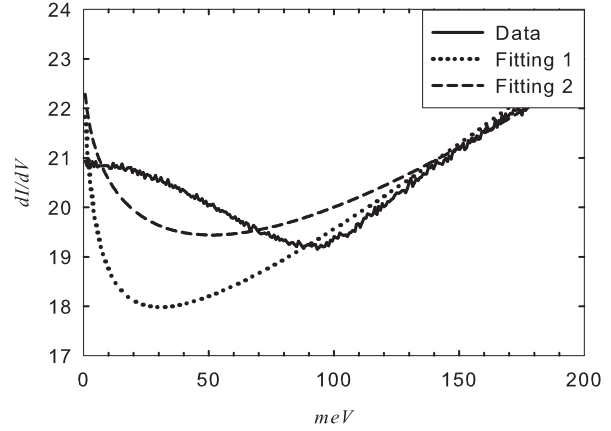}
\caption{\label{fig:fit} Fittings with data taken on BaFe$_2$As$_2$ (solid line) from Ref. \cite{arham2012} with RPA form of Eq. (\ref{fit}) (dotted line) and with
the logarithmic form of Eq. (\ref{logfit}) (dashed line).}
\end{figure}

We can then adopt a general form for the conductance measured in PCS for iron pnictides in which the zero-bias peak has been observed\cite{arham2012}.  The
result
\beq
\frac{dI}{dV} &=& \frac{e^2 T}{h} \big\{D(E_f + B_1 eV) - \frac{C_1 W_1}
{B_1^2 (eV)^2 + W_1^2}\nn\\
&+& \frac{C_1\big[W_1+ W_2 (eV)^\nu\big]}
{\big[B_1 eV - R_1 (eV)^\alpha \big]^2 + \left[W_1 + W_2 (eV)^\nu\right]^2}\big\}
\label{fit}
\eeq
contains the fitting parameters $(B_1,C_1,R_1,W_1,W_2)$. The first term describes the effect from the non-interacting density of states, and $B_1$ captures the effect of the external field 
on the band dispersion which was taken to be $3/2$ in the last
section. The remaining terms describe the effects from the non-Ferimi
liquid self-energy.
The fitting strategy is as follows. Since $(B_1, C_1, W_1)$ are mostly related to the non-interacting part of contribution, they are chosen to fit the PCS conductance at high bias.
$(R_1,W_2)$ are related to the real and imaginary parts of the non-Fermi liquid self energy, and they are fit to obtain the PCS conductance near zero bias.

The fitting to the point-contact conductance for parent BaFe$_2$As$_2$ at $T=T_s=135K$ is shown in Fig. \ref{fig:fit}. The higher bias behavior is fit well by a linear density 
of states ($D(E) = D(E_f) + B_1 (E-E_f)$), which is consistent with STM measurements. As for the zero-bias peak, our theoretical fitting, however, shows a much sharper 
peak at zero bias than is observed experimentally. This deviation
arises from two sources.
First, our treatment is strictly valid at zero temperature. 
At finite temperature, the Fermi function is no longer a step function, so $f'(\epsilon-eV)$ is not a delta function. 
Therefore the convolution of the effective density of states with $f'(\epsilon)$ results in a broadening of the spectral function, directly resulting from thermal population effects.  
To account for this, Eq. (\ref{finalI3}) becomes
\be
I = \frac{e}{\hbar} \int d\vec{p}_\parallel\int^{\infty}_{-\infty} \frac{d\epsilon}{2\pi} T(\vec{p}_\parallel,\epsilon)
A(\vec{p}_\parallel,eV,\epsilon)\big[f(\epsilon-eV) - f(\epsilon)\big],
\label{kbtI}
\ee
where $f(\epsilon) = 1/(e^{\beta\epsilon} + 1)$ and $\beta=1/k_B T$. This results in a
significant modification in the fitting formula ,
\beq
&&\frac{dI}{dV} = \frac{e^2 T}{h} \int \frac{d\epsilon}{2\pi}\big\{D(E_f + B_1 \epsilon) - \frac{C_1 W_1}
{B_1^2 \epsilon^2 + W_1^2}\nn\\
&+& \frac{C_1\big[W_1+ W_2 \epsilon^\nu\big]}
{\big[B_1 \epsilon - R_1 \epsilon^\alpha \big]^2 + \left[W_1 + W_2 \epsilon^\nu\right]^2}\big\}\big[-f'(\epsilon-eV)\big],\nn\\
\eeq
at finite temperature.
This accounts for the fact that the experimentally observed zero-bias peak, performed at finite temperature, is broader than our theory calculated at zero temperature.
Second, we have made the assumption in Eq. (\ref{fit})
that the non-Fermi liquid self-energy is momentum-independent and only
results from a modification of the states right at the Fermi surface. 
In principle, the self-energy should have a smooth transition from a
non-Fermi liquid form to something else for states away from the Fermi surface, and 
unfortunately no current theoretical tools can be employed to
capture this effect accurately. 
Lawler {\it et. al.}  addressed this problem using multidimensional bosonization\cite{lawler2006}, and they found a logarithmic form
for the effective density of states of the form
\be
\int d\vec{p}_\parallel \mathcal{A}(\vec{p}_\parallel,\omega=eV)  = D^*(E_F) + B \omega^{2/3}\ln\omega + \cdots.
\label{logfit}
\ee
As seen in Fig. \ref{fig:fit}, the peak is less sharp if we fit the data with Eq. (\ref{logfit}), as expected.
Finally, at lower temperatures, we measure a sharp dip at zero bias, which always accompanies a background conductance that strongly increases with bias, 
whose origin remains unclear.  
In summary, these details would in general increase the density of states for energies slightly away from the Fermi energy, making the peak at zero bias 
less sharp as seen in experiments.  Except for the sharp dip at zero bias which we cannot explain, we conclude then that any apparent
disagreement with the experimental data can be corrected by finite
temperature and matrix element effects in our formalism.

Consequently, we conclude that a strong zero-bias peak in the PCS
conductance in non-superconducting materials is a genuine feature
of a non-Fermi liquid system. 
Indeed, such a feature represents the fingerprint of strong correlations.  As a comparison, it is well established that PCS can detect quasiparticle scattering 
effects due to electron-phonon, electron-magnon, and single-impurity Kondo interactions, but the feature associated with each is a small ($\sim 1\%$) decrease in the 
background conductance.  The accepted explanation for the small decrease is that, at a particular energy, the electrons are backscattered by the interactions, 
thereby reducing the conductance\cite{jansen1980}.
These features have been successfully modeled using the classical non-linear Boltzmann equation\cite{jansen1980}.  
The only example in which the conductance is increased is in the case of Andreev reflection from superconductors. 
Although this can produce a sub-gap conductance as high as doubling the background conductance, it is due to the unique property of particle-hole mixing of the Cooper pairs\cite{btk1982},
and does not apply in the non-superconducting materials modeled here.  
The experiments we are modeling show a large ($\sim 20\%$) increase in the background conductance around zero bias in non-superconducting materials.  
Unlike the understood effects described just above, we have modeled the quantum-mechanical non-Fermi liquid behavior, resulting from strong electron correlations in which 
the self-energies of all the single-particle states are strongly modified.
The imaginary part of the self-energy determines the broadening of the spectral weight in a particular single-particle state, and the form of ${\rm Im}\Sigma(\epsilon)$ 
defined in Eq. (\ref{selfnfl}) indicates that the states at the Fermi energy suffer the least broadening. As a result, the effective density of states is largest at the Fermi energy,
leading to the zero-bias peak. More importantly, for non-Fermi liquids the power $\nu$ in ${\rm Im}\Sigma(\epsilon) \sim (\epsilon/E_f)^\nu$ is smaller than 1. 
This indicates that ${\rm Im}\Sigma(\epsilon)$ is not negligible even for small $\epsilon$ and consequently, the zero-bias peak is observable in reality.
It is also noteworthy to mention that although the formalism presented here adopts a form obtained from the perturbative RPA approach, the increase of the effective density of states 
at the Fermi energy is also obtained from the non-perturbative multidimensional bosonization technique\cite{lawler2006}. 

Another intriguing feature is that while the DC resistivity shows a significant change with respect to the structural and magnetic phase transitions, the PCS conductance seems to 
be insensitive to them. This discrepancy is due to the different nature of these two measurements. 
DC conductivity, by definition, is the {\it current-current correlation function} at zero frequency $\sigma(0)$, and typically it can be expressed 
using the Drude formula which can be obtained by 
applying the relaxation-time approximation to the Kubo formula\cite{dc}: $\sigma(0) = \frac{ne^2\langle \tau\rangle}{m^*}$,
where $\langle \tau\rangle$ is the transport relaxation time averaged over the Brillouin zone. 
As a result, the DC resistivity is proportional to $1/\langle
\tau\rangle$, which is highly sensitive to the change in the lattice due to the structural phase transition. 
In our formalism, the PCS conductance measures the {\it integrated single-particle spectral function}. 
Since the transport relaxation time is related to the inverse of the self energy
at the Fermi energy, its presence only broadens the delta-function-like peaks in the single-particle spectral function.
As a result, as we integrate the single-particle spectral function over the whole Brillouin zone, Fermi surface reconstructions due to phase transitions play a very small role in the 
measured PCS conductance. 
Exception will be the case when the phase transition opens a gap near the Fermi energy. In this case, the density of states of the reconstructed Fermi surface changes 
radically due to the gap formation, which would be detected by the PCS (see Supplementary Materials).
Since iron pnictides and chalcogenides remains metallic through the structural and magnetic phase transitions, the PCS conductance is insensitive to these phase transitions.

\section{Conclusion}
We have developed a quantum theory to show that PCS is capable of detecting an effective density of states arising from non-Fermi liquid behavior.  
For junctions in the thermal regime, where the electron mean free path is smaller than the junction size, quasiparticle transport is well-described by the current-current 
correlation function according to the Kubo formalism, so the behavior is as a simple resistor.  
As the junction size is reduced to the ballistic (Sharvin) regime, we have shown that the PCS conductance is consistently described with a single-particle Green function, 
according to the SKBK formalism with a set of reasonable approximations.  
We have shown that the microscopic model for PCS can be reduced to a tunneling model with only momentum-conserving processes in the classical (ballistic) limit.  
Furthermore, since PCS samples the effective density of states obtained from integration over whole Fermi surface, it is insensitive to simple Fermi surface reconstruction without a gap, 
so is a filter for non-Fermi liquid-like behavior.  
We have specifically demonstrated that the zero-bias peak observed in the PCS measurement in iron pnictides and chalcogenides is due to a non-Fermi liquid behavior, 
which can be induced by orbital fluctuations.   
This theory can also be used to explain why the non-Fermi liquid density of states can be detected in heavy fermions and quantum critical materials.
Our results shed a new light on application of PCS to investigate strongly correlated systems, and to determine if a new material, in fact, exhibits strong electron correlations. 

\section{Acknowledgement}
We would like to thank Anthony J. Leggett, Jennifer Misuraca, and Mauro Tortello for helpful discussions.
This work is supported by the Center for Emergent Superconductivity, a DOE Energy Frontier Research Center, Grant No. DE-AC0298CH1088. 
WKP is supported by the U.S. NSF DMR under Award 12-06766.

\end{article}

\newpage
\leftline{{\bf Supplementary Materials} }
\section{Non-interacting case}
We first discuss the non-interacting case in which $\Sigma_{sys}=0$. In this case, the only self-energies in the full Green function are from the contacts which can be assumed to be
constants as discussed in previous section. As a result, we have
\be
\Sigma(\vec{p}_\parallel,\epsilon) = \Sigma_L(\vec{p}_\parallel,\epsilon) + \Sigma_R(\vec{p}_\parallel,\epsilon) = \Lambda + i \Gamma,
\ee
where $\Lambda \equiv \Lambda^L + \Lambda^R$ and $\Gamma \equiv \big(\Gamma^L+\Gamma^R\big)/2$.
Plugging the self energy into Eq. (\ref{finalG}), we find that $\mathcal{A}(\vec{p}_\parallel,eV)$ is just a Lorentzian function with the broadening factor $\Gamma$ and a shift in
energy $\Lambda$. Since $\Lambda$ is a constant, it can be absorbed
into the chemical potential which we will drop from now on.
If we take the limit $\Gamma\to 0$, $\mathcal{A}(\vec{p}_\parallel,eV)\to \delta\big(\frac{3}{2} eV - E_{\vec{p}_\parallel}+\mu \big)$, the conductance becomes
\be
\frac{dI}{dV} = \frac{e^2 T}{h} D(\mu + \frac{3}{2} eV ),
\ee
where $D(E)$ is the density of states defined in Eq. \ref{dos}.
This means that the conductance in fact maps out the density of state of the system with energy $\frac{3}{2} eV$ measured from the chemical potential
(or Fermi energy $E_f$ at zero temperature).
It is important to note that in order to obtain the typical Ohmic behavior, $D(E)$ is assumed to be a constant and thus the conductance is a constant as well\cite{cuevas1996}.
This assumption is valid for a metal whose $E_f$ is so large
that the variation of $D(E)$ near Fermi energy is small.
However, for systems with complicated band structure, a non-Ohmic
behavior could be observed if $D(E)$ were to vary significantly near
$E_f$ even in the absence of interaction effects.
In realistic experiments, $\Gamma$ is in general finite, but it will
not produce any qualitative change if it remains small compared to
energy scale we are interested.
Typlically, $\Gamma$ should be no more than a few meVs, otherwise it will be considered as a bad contact\cite{arham2012}.

The result in this section can be applied to some interacting systems in which Fermi surface reconstruction with a gap opening near the Fermi energy occurs at low temperatures and the
associated fluctuations are negligible. 
In this case, the band structure is reconstructed below a certain temperature, and the resulting density of states varies dramatically near the Fermi energy as a 
result of a gap formation\cite{meekes1988}.
Consequently, the conductance measured from the PCS can detect this reconstructed density of states. This may explain the observations made in the PCS study of the Kondo-lattice URu$_2$Si$_2$, in which the reconstruction of the density of states due to the hybridization gap was clearly observed.\cite{park2012}

\section{Itinerant system with unusual self-energy due to strong correlation}
The interaction effects of the system are included in Eq. (\ref{finalG}) through $\Sigma_{sys}$ in the self energy. 
We emphasize that in the derivation of Eq. (\ref{finalG}), only the
contact Hamiltonian $H_c$ in Eq. (\ref{ham}) is treated perturbatively,
to second order in this case. 
We take $G$ in Eq. (\ref{jl}) to be the full Green function. 
In other words, Eq. (\ref{finalG}) is generally valid even for the
spectral function obtained from non-perturbative approaches.
Therefore, just knowing the functional form of the spectral
function, or equivalently $\Sigma_{sys}$, allows us to describe the
features of the conductance.  Quite generally,  larger values of $\Sigma(\vec{p}_\parallel,\epsilon)$ indicate that an electron with momentum $\vec{p}_\parallel$ and 
energy $\epsilon$ undergoes stronger scattering due to interactions.
This effectively reduces the probability for an electron to remain in
its initial state.  
Such a reduction is reflected in the suppression of the spectral function $\mathcal{A}(\vec{p}_\parallel,\epsilon)$. 
The integration of $\mathcal{A}(\vec{p}_\parallel,\epsilon)$ over momentum $\vec{p}_\parallel$ gives the effective density of states at energy $\epsilon$ 
modified by the interaction (self-energy), which is directly proportional to the conductance measured by PCS.

For an itinerant electron system near a quantum critical point, non-Fermi liquid behavior can emerge, and the single-particle Green function is stongly modified. 
While the self-energy can be obtained by either perturbative\cite{oganesyan2001,leewc2012nfl} or non-perturbative approaches\cite{lawler2006, lo2013} for 
specfic models\cite{leewcreview}, it could be written in the following general form
\be
\Sigma_{sys}(\vec{p}_\parallel,\epsilon) = F(\vec{p}_\parallel) V_{eff}\left[a \big(\frac{\epsilon}{E_F}\big)^\alpha +
i \big(b_1 \big(\frac{k_B T}{E_F}\big)^2 + b_2 \big(\frac{\epsilon}{E_F}\big)^\nu\big)\right],
\label{selfnfl}
\ee
where $F(\vec{p}_\parallel)$ is a function peaked at $\vec{p}_\parallel$ on the Fermi surface, $V_{eff}$ the effective interaction parameter, and 
$(a,b_1,b_2)$ are parameters which generally have weak dependence on momenta at the Fermi surface. 
$\epsilon$ here is the energy measured from the Fermi energy.
A standard Fermi liquid has $\alpha=\nu=2$, and the non-Fermi liquid metal is generally defined as $\nu<1$. 
Inserting Eq. (\ref{selfnfl}) into Eq. (\ref{finalG}), we find that the conductance can be divided into non-interacting ($\mathcal{G}_n(eV)$) and interacting ($\mathcal{G}_i(eV)$) 
parts for the simplest case of $F(\vec{p}_\parallel)=1$ for $E_f-\Delta/2\leq E_{\vec{p}_\parallel}\leq E_f+\Delta/2$ and $=0$ elsewhere.
To be more specific, we have
\beq
\frac{dI}{dV} &=& \mathcal{G}_n(eV) + \mathcal{G}_i(eV),\nn\\
\mathcal{G}_n(eV) &=& \frac{e^2 T}{h} D(\mu + \frac{3}{2} eV),\nn\\
\mathcal{G}_i(eV) &=& \frac{e^2 T}{h} \int_{FS} d\vec{p}_\parallel \mathcal{A}(\vec{p}_\parallel,eV),
\eeq
where $\int_{FS}$ refers to integration only over a small area in momentum space near the Fermi surface with $E_f-\Delta/2\leq E_{\vec{p}_\parallel}\leq E_f+\Delta/2$.
 
Care must be taken in analyzing $\mathcal{G}_i(eV)$. First, we introduce $N(E_f) = D(E_f)\Delta$, where $D(E)$ is the non-interacting density of states defined in 
Eq. \ref{dos}, to describe the number of states characterized by
the non-Fermi liquid self-energy. Then $\mathcal{G}_i(eV)$ can be rewritten as
\beq
\mathcal{G}_i(eV) &=& \frac{e^2 T N(E_f)}{h}  \mathcal{A}(E_f,eV) \nn\\
&=& \frac{e^2 T N(E_f)}{E_f h} \frac{\big[\Gamma' + V'_{eff}\big(b_1 T'^2 + b_2 V'^\nu\big)\big]}
{\big[\frac{3}{2}V' - a V'_{eff} V'^\alpha \big]^2 + 
\left[\Gamma' + V'_{eff}\big(b_1 T'^2 + b_2 V'^\nu\big)\right]^2},\nn\\
\label{gint}
\eeq
where we have rescaled the energy with respect to the Fermi energy and introduced the short-hand notation $E'\equiv E/E_f$, $V'\equiv eV/E_f$ and $T'\equiv k_B T/E_f$.
We begin with the case in which the terms indepedent of $eV$ in the imaginary part of self-energy are dropped. 
This is appropriate for a clean junction interface ($\Gamma=0$) at zero-temperature, and Eq. (\ref{gint}) reduces to 
\be
\mathcal{G}^{\Gamma,T=0}_i(eV) = \frac{e^2 T N(E_f)}{E_f h}
\frac{b_2 V'_{eff}}{\frac{9}{4} V'^{p_1} - 3 a V'_{eff} V'^{p_2} + a^2 V'^2_{eff} V'^{p_3} + b_2^2 V'^2_{eff} V'^\nu},
\ee
where 
\be
p_1\equiv 2-\nu,
p_2\equiv \alpha-\nu + 1,
p_3\equiv 2\alpha-\nu,
\ee
are the three powers crucially determining the behavior of $\mathcal{G}^{\Gamma,T=0}_i(eV)$ at $eV=0$. If any of the $p_i$s is smaller than zero, $\mathcal{G}^{\Gamma,T=0}_i(eV=0)=0$. 
If any of $p_i$s is zero and the others are larger than zero, $\mathcal{G}^{\Gamma,T=0}_i(eV=0)$ is a finite value. 
Most importantly, {\it if all the $p_i$s are larger than zero, $\mathcal{G}^{\Gamma,T=0}_i(eV=0)$ diverges}.
As a check, for a Fermi liquid, $\alpha=\nu=2$, and thus $p_1=0$ while $p_2,p_3>0$. As a result, $\mathcal{G}^{\Gamma,T=0}_i(eV=0)$ yields a finite value. 
For a quantum nematic fluid\cite{lawler2006}, $\alpha=\nu=2/3$ so that $p_1,p_2,p_3 >0$. 
Consequently, {\it a divergence of $\mathcal{G}^{\Gamma,T=0}_i(eV=0)$, or a zero-bias peak, obtains}. 
The present theory justifies the interpretation of the zero-bias peak observed in the PCS in a variety of 
iron pnictides\cite{arham2012} in terms of a non-Fermi liquid state
brought on orbital fluctuations,  which was shown\cite{leewc2012nfl,lo2013,leewcreview} to be identical to a quantum nematic fluid.

In the realistic situation, both $\Gamma$ and $T$ are non-zero. However, putting them back to Eq. (\ref{gint}) only renders the behavior for 
$V' < V'_c$, where $V'_c$ is a critical voltage to be determined. 
This will only smooth out the divergence at $V'=0$, and the
conductance will behave similarly to that in $\mathcal{G}^{\Gamma,T=0}_i(eV)$ 
for $V' >  V'_c$. 

The critical voltage $V'_c$ is determined by the condition under
which the terms independent of $V'$ in the imaginary part of self
energy in Eq. (\ref{gint}) can be ignored. 
More specifically, we can estimate $V'_c$ from the condition
\be
\frac{ b_2 V'_{eff} (V'_c)^\nu}{\Gamma' + b_1 V'_{eff}T'^2} \sim 1.
\ee
We unfortunately do not know $\Gamma'$, but it is reasonable to assume
that the junction interface is sufficiently clean so that $\Gamma' << b_1 V'_{eff}(T')^2$.
Then we have
\be
V'_c \sim \big[\frac{b_1 T'^2}{b_2}\big]^{\frac{1}{\nu}}\sim T'^{\frac{2}{\nu}},
\ee
where in the last step we used the fact that $b_1/b_2\sim 1$ for
self-energies given by Eq.(\ref{selfnfl}).
Considering room temperature, $k_B T\sim 25$ meV and the Fermi energy $E_f\sim 2$ eV, which is a reasonable value for most of the materials, t
hen  a non-Fermi liquid emerges with $\nu=2/3$, $V'_c \sim (0.0125)^3$, giving $eV_c\sim (0.0125)^3 E_f = (0.0125)^3 \times 2$ eV $\sim 0.004$ meV.
This estimate shows that the finite temperature effect in the self energy is important at most up to $eV = 0.004$ meV and consequently negligible. 
Therefore, the zero-bias peak should  is a robust feature even at finite temperature.
It is interesting to note that the robustness against the temperature is due to the fact that $\nu < 1$ for a non-Fermi liquid. 
Fermi liquids, for example, have $\nu = 2$, and the critical voltage with the same condition will be $eV_c = (0.0125) \times 2$ eV $=25$ meV, which is much larger 
than the value obtained for the non-Fermi liquids studied here.


\begin{thebibliography}{31}
\bibitem{hfreview1} G. R. Stewart, Non-Fermi-liquid behavior in d- and f-electron metals, Rev. Mod. Phys., 73 (2001), pp. 797-855.
\bibitem{hfreview2} H. v. L\"ohneysen, A. Rosch, M. Vojta, and P. W\"olfle, Fermi-liquid instabilities at magnetic quantum phase transitions, Rev. Mod. Phys., 79 (2007), pp. 1015-1075.
\bibitem{leecupratereview} P. A. Lee, N. Nagaosa, and X.-G. Wen, Doping a Mott insulator: Physics of high-temperature superconductivity, Rev. Mod. Phys., 78 (2006), pp. 17-85.
\bibitem{phillipsreview} P. Phillips, Identifying the propagating charge modes in doped Mott insulators, Rev. Mod. Phys., 82 (2010), pp. 1719-1742.
\bibitem{daireview} J. Dai, Q. Si, J.-X. Zhu, and E. Abrahams, Iron pnictides as a new setting for quantum criticality, Proc. of Nat. Acad. of Sci., 106 (2009), pp. 4118-4121.
\bibitem{leewcreview} W.-C. Lee, W. Lv, and H. Z. Arham, Elementary Excitations due to Orbital Degrees of Freedom in Iron Based Superconductors, Int. J. Mod. Phys. B, 27 (2013) , pp. 1330014.
\bibitem{lawler2006} M. J. Lawler, D. G. Barci, V. Fernandez, E. Fradkin, andnd L. Oxman, Nonperturbative behavior of the quantum phase transition to a nematic Fermi fluid, Phys. Rev. B, 73 (2006), pp. 085101.
\bibitem{caroli} C. Caroli, R. Combescot, P. Nozieres, and D. Saint-James, Direct calculation of the tunneling current, J. Phys. C: Solid State Phys. 4 (1971), pp. 916-929.
\bibitem{tersoff} J. Tersoff and D. R. Hamann, Theory of the scanning tunneling microscope, Phys. Rev. B 31 (1985), pp. 805.
\bibitem{fogelstr2010} M. Fogelstr\"om W. K. Park, L. H. Greene, G. Goll, and M. J. Graf, Point-contact spectroscopy in heavy-fermion superconductors, Phys. Rev. B, 82 (2010), pp. 014527.
\bibitem{yanson1974} I. Yanson, Nonlinear effects in the electric conductivity of point junctions and electron-phonon interaction in normal metals, J. Exp. Theor. Phys., 39 (1974), pp. 506-513.
\bibitem{harrison1961} W. A. Harrison, Tunneling from an Independent-Particle Point of View, Phys. Rev., 123 (1961), pp. 85-89.
\bibitem{jansen1980} A. Jansen, A. van Gelder, and P. Wyder, Point-contact spectroscopy in metals, J. Phys. C: Solid State Phys., 13 (1980), pp. 6073-6118.
\bibitem{khotkevich1995} A. V. Khotkevich and I. K. Yanson, Atlas of Point Contact Spectra of Electron-Phonon Interactions In Metals, Kluwer Academic (1995).
\bibitem{meekes1988} H. Meekes, Point-contact spectroscopy in incommensurate chromium, Phys. Rev. B, 38 (1988), pp. 5924-5930.
\bibitem{escudero2010} R. Escudero and F. Morale, Point contact spectroscopy of Nb$_3$Sn crystals: Evidence of a CDW gap related to the martensitic transition, Solid State Communications, 150 (2010), pp. 715-719.
\bibitem{Zhang2013} X. Zhang, N. P. Butch, P. Syers, S. Ziemak, R. L. Greene, and J. Paglione, Hybridization, Inter-Ion Correlation, and Surface States in the Kondo Insulator SmB$_6$, Phys. Rev. X, 3 (2013), pp. 011011.
\bibitem{btk1982} G. E. Blonder, M. Tinkham, and T. M. Klapwijk, Transition from metallic to tunneling regimes in superconducting microconstrictions: Excess current, charge imbalance, and supercurrent conversion, Phys. Rev. B, 25 (1982), pp. 4515-4532.
\bibitem{park2012} W. K. Park, P. H. Tobash, F. Ronning, E. D. Bauer, J. L. Sarrao, J. D. Thompson, and L. H. Greene, Observation of the Hybridization Gap and Fano Resonance in the Kondo Lattice URu$_2$Si$_2$, Phys. Rev. Lett., 108 (2012), pp. 246403.
\bibitem{goll2005} G. Goll, Point-Contact Spectroscopy on Conventional and Unconventional Superconductors, Adv. Solid State Phys., 45 (2005), 213-225.
\bibitem{park2008} W. K. Park, J. L. Sarrao, J. D. Thompson, and L. H. Greene, Andreev Reflection in Heavy-Fermion Superconductors and Order Parameter Symmetry in CeCoIn$_5$, Phys. Rev. Lett., 100 (2008), pp. 177001.
\bibitem{daghero2011} D. Daghero, M. Tortello, G. Ummarino, and R. Gonnelli, Directional point-contact Andreev-reflection spectroscopy of Fe-based superconductors: Fermi surface topology, gap symmetry, and electron¡Vboson interaction, Rep. Prog. Phys., 74 (2011), pp. 124509.
\bibitem{yang2008} Y.-F. Yang, Z. Fisk, H.-O. Lee, J. D. Thompson, and David Pines, Scaling the Kondo lattice, Nature 454 (2008), pp. 611-613.
\bibitem{haule2009} K. Haule1 and G. Kotliar, Arrested Kondo effect and hidden order in URu$_2$Si$_2$, Nat. Phys. 5 (2009), pp. 796-799.
\bibitem{maltseva2009} M. Maltseva, M. Dzero, and P. Coleman, Electron Cotunneling into a Kondo Lattice, Phys. Rev. Lett. 103 (2009), pp. 206402.
\bibitem{arham2012} H. Z. Arham, C. R. Hunt, W. K. Park, J. Gillett, S. D. Das, S. E. Sebastian, Z. J. Xu, J. S. Wen, Z. W. Lin, Q. Li, et al., Detection of orbital fluctuations above the structural transition temperature in the iron pnictides and chalcogenides, Phys. Rev. B, 85 (2012), pp. 214515; H. Z. Arham, Doctoral Dissertation, University of Illinois (2013).
\bibitem{leewc2012nfl} W.-C. Lee and P. W. Phillips, Non-Fermi liquid due to orbital fluctuations in iron pnictide superconductors, Phys. Rev. B, 86 (2012), pp. 245113.
\bibitem{schwinger} J. Schwinger, Brownian Motion of a Quantum Oscillator, J. Math. Phys. (N.Y.), 2 (1961), pp. 407-432.
\bibitem{kadanoff-baym} L. Kadanoff and G. Baym, Quantum Statistical Mechanics, Benjamin, New York (1962).
\bibitem{lvkeldysh} L. V. Keldysh, Diagram Technique for Nonequilibrium Processes, Sov. Phys. JETP, 20 (1965), pp. 1018-1026.
\bibitem{noting} SKBK formalism was previously applied to PCS to recover BTK results in Andreev scattering: Anders and Gloos, Physica B 230-232 (1997), Towards a microscopic theory for metallic heavy-fermion point contacts, pp. 437-440.
\bibitem{keldysh} H. J. Haug and A.-P. Jauho, Quantum Kinetics in
  Transport and Optics of Semiconductors, Springer (2007); For an
  explicit reference relating the current to the time-ordered Green
  function,
see, D. Dalidovich and P. Phillips, Phys. Rev. Lett. {\bf 93}, 27004 (2004).
\bibitem{cuevas1996} J. C. Cuevas, A. Martin-Rodero, and A. L. Yeyati, Hamiltonian approach to the transport properties of superconducting quantum point contacts, Phys. Rev. B, 54 (1996), pp. 7366-7379.
\bibitem{oganesyan2001} V. Oganesyan, S. A. Kivelson, and E. Fradkin, Quantum theory of a nematic Fermi fluid, Phys. Rev. B, 64 (2001), pp. 195109.
\bibitem{noteAR} Note that this assumption does not hold
for the N/S junction due to the nature of the particle-hole mixing of the Cooper pairs\cite{btk1982}.               
As shown in BTK theory, solving the Bogoliubov equations at the junction automatically leads to momentum-dependent transfer matrix elements for electrons and holes.
This results in the directional PCS conductance induced by Andreev reflection observed in N/S junctions\cite{daghero2011}.
\bibitem{lo2013} K. W. Lo, W.-C. Lee, and P. W. Phillips, Non-Fermi liquid behaviour at the orbital-ordering quantum critical point in the two-orbital model, Europhys. Lett., 101 (2013), pp. 50007.
\bibitem{dc} Y. Murayama, Mesoscopic Systems: Fundamentals and Applications, WILEY-VCH, Verlag Berlin GmbH (2001).

\end{thebibliography}
\end{document}